\documentclass[twocolumn,final]{IEEEtran}
\def\comment#1{}
\usepackage{graphicx,subfigure,subfig,epsfig,amsfonts,amsmath,amssymb,lscape,color,longtable,slashbox}
\usepackage[noadjust]{cite}
\usepackage[mathlines,displaymath]{lineno}
\usepackage{amssymb}
\DeclareSymbolFont{grb}{OML}{cmm}{b}{it}
\DeclareMathSymbol{\zetab}{\mathord}{grb}{"10}
\DeclareMathSymbol{\etab}{\mathord}{grb}{"11}
\DeclareMathSymbol{\thetab}{\mathord}{grb}{"12}
\DeclareMathSymbol{\kappab}{\mathord}{grb}{"14}
\DeclareMathSymbol{\lambdab}{\mathord}{grb}{"15}
\DeclareMathSymbol{\mub}{\mathord}{grb}{"16}
\DeclareMathSymbol{\nub}{\mathord}{grb}{"17}
\DeclareMathSymbol{\rhob}{\mathord}{grb}{"1A}
\DeclareMathSymbol{\sigmab}{\mathord}{grb}{"1B}
\DeclareMathSymbol{\taub}{\mathord}{grb}{"1C}
\DeclareMathSymbol{\phib}{\mathord}{grb}{"1E}
\DeclareMathSymbol{\psib}{\mathord}{grb}{"20}
\DeclareMathSymbol{\omegab}{\mathord}{grb}{"21}
\DeclareMathSymbol{\epsilonb}{\mathord}{grb}{"22}
\DeclareMathSymbol{\varphib}{\mathord}{grb}{"27}

\begin{document}
\bibliographystyle{ieee}

\title{Polynomial-Phase Signal \\Direction-Finding 
\& Source-Tracking \\ with
an Acoustic Vector Sensor}

\author{
\authorblockN{Xin YUAN}\\
\authorblockA{Department of Electrical and Computer Engineering, \\
Duke University, Durham, NC, 27708 \\
xin.yuan@duke.edu}
}


\maketitle
\begin{abstract}
A new ESPRIT-based algorithm is proposed to estimate the direction-of-arrival of an {\em arbitrary} degree polynomial-phase signal with a single acoustic vector sensor.
The proposed approach requires {\em neither} a priori knowledge of the polynomial-phase signal's coefficients {\em nor} a priori knowledge of the polynomial-phase signal's frequency-spectrum.
A pre-processing technique is also proposed to incorporate the single-forgetting-factor algorithm and multiple-forgetting-factor adaptive tracking algorithm to track a polynomial-phase signal using one acoustic vector sensor.
Simulation results verify the efficacy of the proposed direction finding and source tracking algorithms.
\end{abstract}


\section{Introduction}
Direction finding with acoustic vector sensors has attracted much attention
\cite{NehoraiSPT0994,
WongKT_JOE0797,WongKT_JOE1097,HawkesSPT0998,WongKT_SPT1299,WongKT_JOE0400,Zoltowski2SPT0800,TichavskySPT1101,WuIY_SPT0710,ChenSP0405}
in recent years since the acoustic vector sensor outperforms the conventional pressure sensor \cite{NehoraiSPT0994,WongKT_JOE0797,TichavskySPT1101}.
An acoustic vector sensor
comprises three orthogonal
velocity sensors, and a pressure sensor. These four sensors are collocated at a point geometry in space.
The acoustic vector sensor can thus measure both pressure and particle velocity of
the acoustic field at a point in space; whereas a
traditional pressure sensor can only extract the
pressure information.
The response of an acoustic vector sensor to a far-field unity power incident acoustic wave can be characterized by \cite{NehoraiSPT0994}:
\begin{eqnarray}\label{eq:3+1}
{\bf a}
&\stackrel{\rm def}{=}  &
\left[\begin{array}{c}
                {\bf u}_x(\alpha,\beta)\\
                {\bf u}_y(\alpha,\beta)\\
                {\bf u}_z(\alpha) \\
                1  \end{array}\right]
 \hspace{0.08in}\stackrel{\rm def}{=}  \hspace{0.08in}
\left[\begin{array}{c}
                \sin\alpha \hspace{0.01in} \cos\beta\\
                \sin\alpha \hspace{0.01in} \sin\beta\\
                \cos\alpha  \\
                1 \end{array}\right],
\end{eqnarray}
where $\alpha\in [0,\pi],\beta\in[0,2\pi)$ are the elevation-angle and azimuth-angle of the source, respectively, and ${\bf u}_x, {\bf u}_y, {\bf u}_z$ symbolize the three Cartesian components of ${\bf u}$ along each axis in the Cartesian coordinate system, respectively.
Much work has been done by applying the acoustic vector sensor for direction-finding \cite{NehoraiSPT0994,
WongKT_JOE0797,WongKT_JOE1097,HawkesSPT0998,HawkesJOE0199,
WongKT_SPT1299,WongKT_JOE0400,Zoltowski2SPT0800,TichavskySPT1101,
HawkesSPT0603,XuSP0707,HeIET-RSN0609,TamSJ0809,WongKT_AEST0110,HeJOE0110,LiSPT0111,WuIY_SPT0710,ChenSP0405}, sensor modeling \cite{AbdiSPT0309}, beampattern \cite{WongKT_JOE0702} and beamforming \cite{ChenPRSN0604,LockwoodJASA0106,ZouSPT0809}.
Many advantages are offered by this acoustic vector sensor \cite{WuIY_SPT0710}: a) The array-manifold is independent of the source's frequency spectrum.
b) The array-manifold is less sensitive to the distance of the source.
However, overlooked in the literature is how to estimate the direction-of-arrival of a polynomial-phase signal with arbitrary degree.

Polynomial-phase signal (PPS) is a model used in a variety of applications. For example:
radar, sonar, and communication systems use continuous-phase
modulation where the amplitude is constant and the
phase is a continuous function of time \cite{AdjradSPT0107}.
This function on a closed interval can be uniformly approximated by polynomials from the Weierstrass theorem \cite{Phillips03}.
The phase of the signal above can then be modeled as a
finite-order polynomial within a finite-duration time-interval.
A unity power polynomial-phase signal can be modeled in continuous time as:
\begin{eqnarray}
s(t) &=& \exp\left\{j\left(b_0+ b_1 t + b_2 t^2 + \cdots + b_q t^q\right)\right\},  \label{eq:stpsib}
\end{eqnarray}
where $b_0,b_1, \cdots, b_q$ are the polynomial coefficients associated with the corresponding orders, $q$ is the degree of the polynomial-phase signal, and the initial phase is $b_0$.
When $q=2$, the polynomial-phase signal is known as an LFM (linear frequency modulated) signal.
The polynomial-phase signal has received considerable attention
in the literature
\cite{
PelegITT0391,
FrancosSPT0799,
BenidirSPT0799,
FarquharsonSPT0805,
PhamSPT0107,
PhamSPT0408,
WangSPT0708,
WuSPT1008,
WangSPL0909,
McKilliamSPT1109,
WangSPT0410,
O'sheaAEST0710}.
During the last decade, there has been a growing interest in estimating the parameters
of polynomial-phase signal impinging on a
multi-sensor array
\cite{GershmanSPT1201,
MaSPT0506,
AdjradSPT0107,
AmarSPT0210,
AmarSPT0910}.
Even though both the acoustic vector sensor and the polynomial-phase signal have been extensively investigated during the past two decades, how to estimate the azimuth-elevation angle of a polynomial-phase signal with an acoustic vector sensor seems to be overlooked in the literature.
The present paper fills this gap by proposing an ESPRIT-based algorithm to estimate the direction-of-arrival (DOA) of an arbitrarily deterministic degree polynomial-phase signal.
Given the degree of the polynomial-phase signal, this approach requires {\em neither} a priori knowledge of the polynomial-phase signal's coefficients {\em nor} a priori knowledge of the polynomial-phase signal's frequency-spectrum.
Furthermore, a pre-processing technique is also proposed to adopt the single-forgetting-factor tracking algorithm and the multiple-forgetting-factor tracking algorithm to improve the tracking performance when the acoustic vector sensor is used to track a polynomial phase signal.

\section{Proposed Algorithm -- Direction Finding}
\label{Sec:AL}
\subsection{Measurement Model}
\label{Sec:MM}

Consider a polynomial-phase signal impinging upon an acoustic vector sensor.
The collected $4 \times 1$ data vector
at time $t$ equals:
\begin{eqnarray} \label{eq:zt}
{{\bf z}}(t) &=& {\bf a} s(t) + {\bf n}(t),
\end{eqnarray}
where ${\bf n}(t)$ symbolizes the additive noise at the acoustic vector sensor, $s(t)$ is the polynomial-phase signal as in (\ref{eq:stpsib}),
${\bf a}$ is the steering vector of the signal as in (\ref{eq:3+1}).
In this work, ${\bf n}(t)$ is modeled as zero mean, Gaussian distributed, and with a covariance of a $4\times4$ diagonal matrix ${\bf K}_0= {\rm diag}[\sigma^2,\sigma^2,\sigma^2,\sigma^2]$, where $\sigma^2$ denotes the variance of noise
collected by each constituent antenna.
\footnote{This proposed algorithm can also be used in the three-component acoustic vector sensor as discussed in \cite{WongKT_JOE0400}.}

\subsection{Derivation of the Matrix-Pencil Pair}
\label{Sec:ALMP}

For an acoustic vector sensor, from (\ref{eq:zt}):
\begin{eqnarray}
{\bf z}(t) &=& {\bf a} s(t) + {\bf n}(t) \notag\\
&=& \left[
{\bf u}_x, \hspace{0.05in} {\bf u}_y, \hspace{0.05in} {\bf u}_z,\hspace{0.05in}
1\right]^T s(t)+{\bf n}(t),
\end{eqnarray}
where $^T$ denotes the transposition.
In order to simplify the exposition, we consider the {\em noiseless} case in the following derivation.
Consider a $q$-order polynomial-phase signal, and let ${\bf z}_{(q)}(t)$ be the measured data of the acoustic vector sensor for this signal. In
the noiseless case:
\begin{eqnarray}
{\bf z}_{(q)}(t) &=& \left[
{\bf u}_x, \hspace{0.05in} {\bf u}_y, \hspace{0.05in} {\bf u}_z,\hspace{0.05in}
1\right]^T s(t).
\end{eqnarray}
${\bf z}_{(q)}(t)$ is a $4 \times 1$ vector, and let ${z}_{i,q}(t)$ be the $i$th row of  ${\bf z}_{(q)}(t)$, $\forall i=1,2,3,4$.
With $\delta_T$ to denote a constant time-delay, when $q\ge2$, perform the following computation:
\begin{enumerate}
\item[1)]
For any $\delta_T \neq 0$,
\begin{eqnarray}
{\bf z}_{(q-1)}(t) &\stackrel{\rm def}{=} & {\bf z}_{(q)}(t) {z}^*_{i,q}(t+ \delta_T), \label{eq:xqt}
\end{eqnarray}
where $^*$ denotes the complex conjugation.
\item[2)]
Repeat step 1) for $q = q-1$ until ${\bf z}_{(1)}(t)$ is reached.
\end{enumerate}
For a $q$-order PPS, in total there are $(q-1)$ times recursive computation of step 1).
\footnote{When $q=1$, the frequency of the PPS is a constant, and the PPS is thus a pure-tone. In this case, the
proposed algorithm will degenerate to the ``univector hydrophone ESPRIT" algorithm in \cite{TichavskySPT1101}. It will require
no recursive computation of steps 1), and can be used in the multiple-source
scenario directly. For details, please refer to \cite{TichavskySPT1101}.}

It is known that for every recursive computation of step 1), one-order difference-function of the signal's phase is derived.
Since the phase of the  $q$-order PPS is a $q$-order polynomial of $t$, the $(q-1)$-order difference-function is a $1$-order polynomial.
Thus, ${\bf z}_{(1)}(t)$ is the $1$-order polynomial of $t$. With some manipulation:
\begin{eqnarray}
{\bf z}_{(1)}(t) &=& {\bf a}\left( \left|[{\bf a}]_i\right|^{\left(2^{(q-2)}-1\right)} [{\bf a}]^*_i\right) \notag\\
 && \cdot e^{j(-1)^{(q-1)}\left[f(\delta_T, b_{q-1}, b_q) + (q!)b_q \delta^{(q-1)}_T t\right]} \notag\\
&=& \underbrace{{\bf a}\left( \left|[{\bf a}]_i\right|^{\left(2^{(q-2)}-1\right)} [{\bf a}]^*_i\right) \cdot e^{j(-1)^{(q-1)}f(\delta_T,  b_{q-1}, b_q)}}_{\stackrel{\rm def}{=} {\tilde {\bf a}}} \notag\\
&&\cdot e^{j(-1)^{(q-1)}(q!)b_q \delta^{(q-1)}_T t},\hspace{0.2in}\forall q\geq2; \label{eq:x1t}
\end{eqnarray}
where $[{\bf a}]_i$ denotes the $i$th element in ${\bf a}$, $\left|[{\bf a}]_i\right|$ is the absolute value of $[{\bf a}]_i$, $q !=1\times 2\times 3\times \cdots\times q$ refers to the factorial of $q$, and $f(\delta_T, b_{q-1}, b_q)$ is a function of the parameters in the $(\hspace{0.03in})$. Note that $f(\delta_T, b_{q-1}, b_q)$ is independent of $t$, and for different $q$, it has different values.

Introducing another constant time-delay ${\Delta_T}$:
\begin{eqnarray}
{\bf z}_{(1)}(t) &=& {\tilde {\bf a}}e^{j(-1)^{(q-1)}(q!)b_q \delta^{(q-1)}_T t}, \\
{\bf z}_{(1)}(t+ \Delta_T) &=& {\tilde {\bf a}} e^{j(-1)^{(q-1)}(q!)b_q \delta^{(q-1)}_T (t+\Delta_T)} \notag\\
 &=&   {\bf z}_{(1)}(t) e^{j(-1)^{(q-1)}(q!)b_q \delta^{(q-1)}_T \Delta_T}.
\end{eqnarray}
In practical applications, $\Delta_T$ can be the same as or different from $\delta_T$.

The entire $8 \times 1$ data set is:
\begin{eqnarray} \label{eq:y1y2}
{\bf y} &\stackrel{\rm def}{=}& \left[\begin{array}{l}{\bf z}_{(1)}(t) \\
{\bf z}_{(1)}(t + \Delta_T) \end{array}\right]\notag\\
&\stackrel {\rm def}{=}& \left[\begin{array}{c}
{\bf y}_1\\
{\bf y}_2\end{array}\right] = \left[\begin{array}{l}
{\bf y}_1 \\
{\bf y}_1 e^{j (-1)^{(q-1)} b_q (q!) \delta^{(q-1)}_T \Delta_T}\end{array}\right].
\end{eqnarray}

Note that $e^{j (-1)^{(q-1)} b_q (q!) \delta^{(q-1)}_T \Delta_T}$ depends on (i) the highest-order polynomial-coefficient $b_q$, (ii) the degree of the polynomial-phase signal $q$, and (iii) the time-delays $\{\delta_T, \Delta_T\}$, all of which are {\em constants}. Thus, $e^{j (-1)^{(q-1)} b_q (q!) \delta^{(q-1)}_T \Delta_T}$ is {\em time-independent} and will be used as the {\em invariant-factor} in the following ESPRIT \cite{RoyASSPT0789} algorithm.

Suppose there are $N$ snapshots collected in $\{{\bf z}_{(1)}(t), {\bf z}_{(1)}(t+ \Delta_T)\}$. Then construct the $8 \times N$ data set:
\begin{eqnarray}
{\bf Y} &\stackrel{\rm def}{=}& \left[{\bf y}(t_1), {\bf y}(t_2),\cdots, {\bf y}(t_N)\right] = \left[\begin{array}{c}
{\bf Y}_1 \\ {\bf Y}_2 \end{array}\right].\label{eq:bfY}
\end{eqnarray}

{\bf{\em Remarks:}}
\begin{enumerate}
\item[$\bullet$] In (\ref{eq:xqt}), step 1) to compute the ${\bf z}_{(1)}(t)$, any one row in ${\bf z}_{(q)}(t)$ can be used. This does not affect the following derivation. In cases when any one row is equal to zero, we can use any other nonzero entity.
\item[$\bullet$](\ref{eq:xqt}) in step 1) can be changed to:
\begin{eqnarray}
{\bf z}_{(q-1)}(t) &=& \sum^4_{i=1}{\bf z}_{(q)}(t) z^*_{i,q}(t+ \delta_T). \label{eq:sumxqt}
\end{eqnarray}
Though (\ref{eq:sumxqt}) will increase the computation workload, it has the following advantages:
a) preserving the signal, b) enhancing the noise cancelation, and c) avoiding the case when one row in ${\bf z}_{(q)}(t)$ is equal to zero.
\item[$\bullet$] Equation (\ref{eq:x1t}) holds in the single-source scenario and also for the algorithm derived in this section. In the multiple-source scenario, the algorithm to separate the source-of-interest should first be used and the proposed algorithm can then be adopted in a single-source scenario.
\item[$\bullet$] In the {\em noisy} case, {\em multiplicative noise} will be introduced in (\ref{eq:xqt}). Equation (\ref{eq:x1t}) will become approximated.
    When the noise power $\sigma^2$ increases, the noise will affect the algorithm adversely. With the fixed PPS at the deterministic DOA, when the degree of the PPS increases, the repetitions of step 1) will increase. Thus more {\em multiplicative noise} will be introduced, which will affect the algorithm more seriously.
\end{enumerate}

\subsection{Adopting ESPRIT to Above Data-sets}

The data set ${\bf Y}$ in (\ref{eq:bfY}) can be seen as a data vector based on the vector $\tilde {\bf a}$ defined in equation (\ref{eq:x1t}) (which is modified from the array-manifold $\bf a$).
Compute the correlation matrix of the $8\times N$ data measurements:
\begin{align}
{\bf Y}{\bf Y}^H&= \left[\begin{array}{c}
{\bf Y}_1 \\
{\bf Y}_2 \end{array}\right] \left[\begin{array}{cc}
{\bf Y}^H_1 &{\bf Y}^H_2
 \end{array}\right]=\left[\begin{array}{cc}
{\bf Y}_1{\bf Y}^H_1 &{\bf Y}_1{\bf Y}^H_2 \\
{\bf Y}_2{\bf Y}^H_1 &{\bf Y}_2{\bf Y}^H_2 \end{array}\right],
\end{align}
and then carry on the eigen-decomposition, where $^H$ denotes conjugate transposition.

Similar to Section III-B in \cite{TichavskySPT1101}, there are two estimates of the steering vector $\hat{\bf v}_1$ (corresponding to ${\bf Y}_1{\bf Y}^H_1$), $\hat{\bf v}_2$ (corresponding to ${\bf Y}_2{\bf Y}^H_2$). Since in the present work, we only consider the one-source scenario, these two estimates are obtained from the eigenvector of  ${\bf Y}{\bf Y}^H $ associated with the
largest eigenvalue ($\hat{\bf v}_1$ corresponds to the top $4\times 1$ sub-vector, and $\hat{\bf v}_2$ corresponds to the bottom $4\times 1$ sub-vector). They are inter-related by the value ${\rho}=e^{j(-1)^{(q-1)}(q!)b_q \delta^{(q-1)}_T \Delta_T}$, and this ${\rho}$ can be estimated by the two
estimated steering vectors $\hat{\bf v}_1, \hat{\bf v}_2$ through:
\begin{eqnarray}
\hat{\rho}=(\hat{\bf v}_1^H\hat{\bf v}_1)^{-1}\hat{\bf v}_1^H\hat{\bf v}_2. \label{eq:rho}
\end{eqnarray}
\footnote{the $q$th-order polynomial coefficient can be estimated by:
$\hat{b}_q = \frac{\angle{\hat \rho} + 2\pi m_b}{(-1)^{(q-1)}(q!)\delta^{(q-1)}_T \Delta_T},$
where $m_b$ is an integer and can be determined by a priori knowledge of the region of $b_q$.
After the estimation of DOA, the other polynomial coefficients can be estimated from the algorithms derived in the corresponding references.}
Therefore, $\tilde {\bf a}$ can be estimated from $\hat{\bf v}_1, \hat{\bf v}_2$ by (within an unknown complex number $c$):
\begin{eqnarray}
\hat{\tilde {\bf a}} &=& \frac{1}{2}\left(\hat{\bf v}_1 + \frac{\hat{\bf v}_2}{\hat {{\rho}}}\right) = c {\tilde {\bf a}}. \label{eq:hata}
\end{eqnarray}

It is worth noting that the algorithm can be used for an {\em arbitrary} degree polynomial-phase signal
(i.e, if $q=2$, it is an LFM signal). Given the degree of the polynomial-phase signal, the algorithm requires {\em no} a priori knowledge of the polynomial coefficients.
Since the derivation of the matrix-pencil pair depends solely on the degree of the PPS, the efficacy of the proposed algorithm is {\em independent} of the polynomial coefficients of the signal.

It follows that:
\begin{eqnarray}
\hat{\bf u}_x &=& \frac{[{\hat{\tilde {\bf a}}}]_1}{[{\hat{\tilde {\bf a}}}]_4},\hspace{0.1in}
\hat{\bf u}_y \hspace{0.07in} = \hspace{0.07in} \frac{[{\hat{\tilde {\bf a}}}]_2}{[{\hat{\tilde {\bf a}}}]_4}, \hspace{0.1in}
\hat{\bf u}_z \hspace{0.07in}= \hspace{0.07in} \frac{[{\hat{\tilde {\bf a}}}]_3}{[{\hat{\tilde {\bf a}}}]_4}.
\end{eqnarray}

Lastly, the direction-of-arrival of the polynomial-phase signal can be estimated by:
\begin{eqnarray}
{\hat \alpha} &=&  \arccos\left({\hat {\bf u}_z}\right) ,\hspace{0.1in}
{\hat \beta} \hspace{0.07in}=\hspace{0.07in} \angle\left(\hat{\bf u}_x + j{\hat{\bf u}_y}\right),
\end{eqnarray}
where $\angle$ denotes the angle of the following complex number.

\section{Proposed Algorithm -- Source Tracking}
\label{Sec:ST}

If the source is moving, the DOA of the source will become time-varying and the array-manifold will change with the time. Therefore (\ref{eq:3+1}) will become:
\begin{eqnarray}
{\bf a}(\alpha(t),\beta(t))
&\stackrel{\rm def}{=}  &
\left[\begin{array}{c}
                {\bf u}_x(\alpha(t),\beta(t))\\
                {\bf u}_y(\alpha(t),\beta(t))\\
                {\bf u}_z(\alpha(t)) \\
                1  \end{array}\right] \notag\\
&\stackrel{\rm def}{=} &
\left[\begin{array}{c}
                \sin(\alpha(t)) \hspace{0.01in} \cos(\beta(t))\\
                \sin(\alpha(t)) \hspace{0.01in} \sin(\beta(t))\\
                \cos(\alpha(t)) \\
                1 \end{array}\right].
\end{eqnarray}

With $T_s$ to denote the sampling time interval, consider there are $M$ time samples. The collected $4\times M$ data set will be:
\begin{eqnarray}
{\bf Z} &=& \left[{\bf z}(T_s), {\bf z}(2T_s), \cdots, {\bf z}(MT_s)\right], \label{eq:Z}
\end{eqnarray}
where ${\bf z}(mT_s)= {\bf a}(\alpha(mT_s),\beta(mT_s)) s(mT_s) + {\bf n}(mT_s)$, $\forall m=1,2,\cdots, M$.

Consider the first $(q+1)$ data vectors in (\ref{eq:Z}), $[{\bf z}(T_s), {\bf z}(2T_s), \cdots, {\bf z}((q+1)T_s)]$.
Recall the pre-processing steps of the date sets in Section \ref{Sec:ALMP} by setting $\delta_T=T_s$, and presume the elevation-azimuth angle of the source remains the same during the time interval $(q+1)T_s$. For a $q$-order polynomial-phase signal,
perform one more computation of step 1) from the above $(q+1)$  data vectors, and in total, there will be $q$ times computation. The following result will be obtained:
\begin{eqnarray}
\breve{{\bf z}}(T_s)&=& \breve{\bf a}(\alpha(T_s), \beta(T_s)) e^{j (-1)^{q} b_q (q!) T_s^q}, \\
\breve{\bf a} &\stackrel{\rm def}{=}& {\bf a}\left( \left|[{\bf a}]_i\right|^{\left(2^{(q-2)}\right)} [{\bf a}]^*_i\right).
\end{eqnarray}
Similarly, from any $(q+1)$ contiguous data vectors in (\ref{eq:Z}), $[{\bf z}(nT_s), {\bf z}((n+1)T_s), \cdots, {\bf z}((n+q)T_s)]$, we can obtain:
\begin{eqnarray}
\breve{{\bf z}}(nT_s)&=& \breve{\bf a}(\alpha(nT_s), \beta(nT_s)) e^{j (-1)^{q} b_q (q!) T_s^q}, \notag\\
&& \hspace{0.2in}\forall n=1,2, \cdots, (M-q).
\end{eqnarray}

The following problem is to adaptively estimate $(\alpha(nT_s),\beta(nT_s))$ over $n=1,2,\cdots,(M-q)$, from $[\breve{{\bf z}}(T_s),\breve{{\bf z}}(2T_s),\cdots,\breve{{\bf z}}((M-q)T_s)]$.
The algorithms in \cite{NehoraiSPT0994,NehoraiSPT1099,WongKT_OCEANS-AP06} can be adopted for the source-tracking of the polynomial-phase signal.
The above manipulations extract the relation among the $(q+1)$ adjacent data sets for the polynomial-phase signal, the estimate based on
$[\breve{{\bf z}}(T_s),\breve{{\bf z}}(2T_s),\cdots,\breve{{\bf z}}((M-q)T_s]$ will thus outperform the estimate from $[{\bf z}(T_s), {\bf z}(2T_s), \cdots, {\bf z}(MT_s)]$ directly. The simulation results in Section \ref{Sec:simm} verify this point.
The following reviews the ``single-forgetting-factor tracking"
\footnote{For the Multiple-Forgetting-Factor (MFF) tracking approach in \cite{NehoraiSPT1099,WongKT_OCEANS-AP06}, the described pre-processing technique can also be adopted.}
algorithm in \cite{NehoraiSPT1099,WongKT_OCEANS-AP06}.

\subsection{Review the Single-Forgetting-Factor Tracking Algorithm in \cite{NehoraiSPT1099,WongKT_OCEANS-AP06}}
The recursive least-squares algorithm is used for the source-tracking in \cite{NehoraiSPT1099,WongKT_OCEANS-AP06} as:
\begin{eqnarray}
{\hat{\breve {\bf a}}}(n T_s)&=& \frac{{\mathfrak Re}\left\{\breve{{\bf z}}(nT_s) \right\}}{{\mathfrak Re}\left\{[\breve{{\bf z}}(nT_s)]_4\right\}}, \notag\\
{\hat {\bf a}}_N&=&\left[\begin{array}{c}{\hat{\bf u}}_{x,N}\\
{\hat{\bf u}}_{y,N}\\
{\hat{\bf u}}_{z,N} \\
1\end{array}\right] \hspace{0.07in}=\hspace{0.07in}\frac{\sum^{N}_{n=0}\lambda^{-n} {\hat{\breve {\bf a}}}(n T_s)}{\sum^{N}_{n=0}\lambda^{-n}}, \notag
\end{eqnarray}
where $\lambda<1$ denotes a ``forgetting factor" and  ${\mathfrak Re}\{.\}$ denotes the real-value part of the entity inside $\{.\}$,
It follows that the recursive relation is obtained:
\begin{eqnarray}
{\hat {\bf a}}(nT_s) = \lambda {\hat {\bf a}}(nT_s-T_s) + (1-\lambda){\hat{\breve {\bf a}}}(n T_s),  \forall n=1,2,\cdots, N. \notag
\end{eqnarray}
Hence,
\begin{eqnarray}
{\hat \alpha}_N &=& \arccos \left({\hat{\bf u}}_{z,N}\right), \hspace{0.1in}
{\hat \beta}_N \hspace{0.07in}=\hspace{0.07in} \angle \left({\hat{\bf u}}_{x,N} + j {\hat{\bf u}}_{y,N} \right). \notag
\end{eqnarray}

\section{Cram\'{e}r-Rao Bounds Derivation}

Cram\'{e}r-Rao bound (CRB) is an essential benchmark used to evaluate the performance of various unbiased estimators.
Different Cram\'{e}r-Rao bounds (CRBs) for the acoustic vector sensor have been derived in \cite{NehoraiSPT0994,TamSJ0809,Ahmadi-ShokouhSPT0607,WuIY_SPT0710}.
The Gaussian signal model was used in \cite{NehoraiSPT0994} and closed-form CRBs in a single-source single-vector-sensor scenario were presented in \cite{NehoraiSPT0994}.
\cite{Ahmadi-ShokouhSPT0607} aimed to find the acoustic vector-sensor's minimal composition for finite estimation-variance in direction-finding.
The three collocated velocity-sensors were recommended for boundaryless direction-finding, while a pressure-sensor collocated with the $x$-axis and $y$-axis velocity-sensors was recommended for direction-finding near a boundary. Only the Fisher Information Matrix was derived in \cite{Ahmadi-ShokouhSPT0607}.
Different from the Cram\'{e}r-Rao bounds derived in \cite{NehoraiSPT0994,Ahmadi-ShokouhSPT0607}, the CRBs under non-ideal gain-phase responses, non-collocation, or non-orthogonal orientation were derived in \cite{TamSJ0809}.  The signal model used in \cite{TamSJ0809} was a pure tone incident source.

Unlike the studies above, this paper will derive new Cram\'{e}r-Rao bounds for the acoustic vector sensor in a polynomial-phase signal scenario.
Like the previous studies, the additive complex Gaussian-distributed
noise model will also be used in the following derivation.

\subsection{The Statistical Data Model}
\label{Sec:SDM}
Recall the measurement model in (\ref{eq:zt}), and let $\kappab=[\alpha,\beta,b_0,b_1,\cdots,b_q]^T$ collect all the unknown parameters. The noise covariance $\sigma^2$ is modeled as a priori known.
With $N$ number of time samples, the collected $4N \times 1$ data set equals:
\begin{eqnarray}
{\bf v} &\stackrel{\rm def}{=}& \left[ {\bf z}^T(T_s), {\bf z}^T(2T_s), \cdots, {\bf z}^T(NT_s)\right]^T \notag\\
&\stackrel{\rm def}{=}& {\bf m}(\kappab) + {\bf w},
\end{eqnarray}
where ${\bf m} (\kappab) \stackrel{\rm def}{=} {\bf a} \otimes {\bf s}$ with
${\bf s} \stackrel{\rm def}{=} \left[s(T_s), s(2T_s), \cdots, s(N T_s)\right]^T$,
${\bf w} \sim   {\cal N}(0, \sigma^2 {\bf I}_{4M})$
denotes a zero-mean, Gaussian distributed process, with a covariance matrix ${\bf K}=\sigma^2 {\bf I}_{4M}$,
${\bf I}_{4M}$ is a $4M \times 4M$ identity matrix, and $\otimes$ symbolizes the Kronecker product.
It follows that ${\bf v}\sim   {\cal N}({\bf m} (\kappab), {\bf K}) $.

\subsection{Deriving the Cram\'{e}r-Rao Bounds for Direction Finding}
In the statistical data model in Section \ref{Sec:SDM}, the $(q+3)$ unknown parameters in $\kappab$ introduce a $(q+3)\times (q+3)$ Fisher Information Matrix (FIM):
\begin{eqnarray}
{\bf J} =& \left[\begin{array}{cccccc}
{J}_{\alpha,\alpha} & {J}_{\alpha,\beta}  & {J}_{\alpha,b_0} & {J}_{\alpha,b_1} & \cdots & {J}_{\alpha,b_q}   \\
{J}_{\beta,\alpha} & {J}_{\beta,\beta} & {J}_{\beta,b_0} & {J}_{\beta,b_1} & \cdots & {J}_{\beta,b_q}   \\
{J}_{b_0,\alpha} & {J}_{b_0,\beta}  & {J}_{b_0,b_0} & {J}_{b_0,b_1} & \cdots & {J}_{b_0,b_q}   \\
{J}_{b_1,\alpha} & {J}_{b_1,\beta}  & {J}_{b_1,b_0} & {J}_{b_1,b_1} & \cdots & {J}_{b_1,b_q}   \\
\vdots & \vdots & \vdots & \vdots  &\ddots &\vdots \\
{J}_{b_q,\alpha} & {J}_{b_q,\beta}  & {J}_{b_q,b_0} &  {J}_{b_q,b_1} &\cdots & {J}_{b_q,b_q}   \\
\end{array}\right].
\end{eqnarray}
As all the parameters are independent of ${\bf K}$, from equation (8.34) in \cite{VanTrees02} by setting the second term to zero, the $(i,j)$th entry of  $\bf J$ is:
\begin{eqnarray}
[{\bf J}]_{i,j}
&=&
2{\mathfrak Re} \left\{\frac{\partial{\bf m}^H(\kappab)}{\partial [\kappab]_i}
{\bf K}^{-1}
                \frac{\partial {\bf m}(\kappab)}{\partial [\kappab]_j }\right\}.
 \label{eq:Jij}
\end{eqnarray}

The Cram\'{e}r-Rao bounds of the direction-of-arrival are:
\begin{eqnarray}
\mbox{CRB}(\alpha)   &=&  \left[{\bf J}^{-1} \right]_{1,1}, \nonumber\\
\mbox{CRB}(\beta) &=&  \left[{\bf J}^{-1} \right]_{2,2}.
\end{eqnarray}

The following will show the intermediate steps to derive the elements in the Fisher Information Matrix.
\begin{eqnarray}
\frac{\partial {\bf m}}{\partial \alpha} &=& \frac{\partial {\bf a}}{\partial \alpha} \otimes {\bf s}, \notag\\
\frac{\partial {\bf m}}{\partial \beta} &=& \frac{\partial {\bf a}}{\partial \beta} \otimes {\bf s}, \notag\\
\frac{\partial {\bf m}}{\partial b_{\ell}} &=& \frac{\partial {\bf s}}{\partial b_{\ell}} \otimes {\bf a}, \hspace{0.2in} \ell= 0,1,\cdots q. \notag\\
\frac{\partial {\bf a}}{\partial \alpha} &=& \left[\cos\alpha\cos\beta, \cos\alpha\sin\beta,-\sin\alpha,0\right]^T,\notag\\
\frac{\partial {\bf a}}{\partial \beta} &=& \left[-\sin\alpha\sin\beta, \sin\alpha\cos\beta,0,0\right]^T,\notag\\
\frac{\partial s (n T_s)}{\partial b_{\ell}} &=& j (nT_s)^{\ell} s (n T_s), \hspace{0.3in} n=1,2,\cdots, N. \notag\\
\frac{\partial {\bf s}}{\partial b_{\ell}} &=& \left[j (T_s)^{\ell} s (T_s), j (2T_s)^{\ell} s (2T_s),\right.\notag\\
&& \hspace{0.5in} \left.\cdots, j (N T_s)^{\ell} s (NT_s)\right]^T. \notag
\end{eqnarray}

\begin{eqnarray}
J_{\alpha,\alpha} &=& 2{\mathfrak Re} \left[\left(\frac{\partial {\bf m}}{\partial \alpha} \right)^H
{\bf K}^{-1}
                \left(\frac{\partial {\bf m}}{\partial \alpha }\right)\right]\notag\\
                &=& 2{\mathfrak Re} \left[\left(\frac{\partial {\bf m}}{\partial \alpha}  \otimes {\bf s}\right)^H
{\bf K}^{-1}
                \left(\frac{\partial {\bf m}}{\partial \alpha}  \otimes {\bf s}\right)\right] \notag\\
                &=& \frac{2N}{\sigma^2}\left[\left(\frac{\partial {\bf m}}{\partial \alpha} \right)^H\frac{\partial {\bf m}}{\partial \alpha} \right]
                \notag\\
                &=& \frac{2N}{\sigma^2};  \notag\\
J_{\alpha,\beta} &=& J_{\beta,\alpha} =     \frac{2N}{\sigma^2}\left[\left(\frac{\partial {\bf m}}{\partial \alpha} \right)^H\frac{\partial {\bf m}}{\partial \beta} \right] \notag\\
                &=& 0;  \notag\\
J_{\beta,\beta} &=&     \frac{2N}{\sigma^2}\left[\left(\frac{\partial {\bf m}}{\partial \beta} \right)^H\frac{\partial {\bf m}}{\partial \beta} \right] = \frac{2N \sin^2\alpha}{\sigma^2};  \notag \\
J_{\alpha,b_{\ell}} &=&   J_{b_{\ell},\alpha} = 2{\mathfrak Re} \left[\left(\frac{\partial {\bf a}}{\partial \alpha}  \otimes {\bf s}\right)^H
{\bf K}^{-1}
                \left(\frac{\partial {\bf s}}{\partial b_{\ell}} \otimes {\bf a}\right)\right] \notag \\
                &=&\frac{2{\mathfrak Re} \left[j \left(\frac{\partial {\bf a}}{\partial \alpha}\right)^H {\bf a}\right]}{\sigma^2} T^{\ell}_s\sum^N_{n=1} n^{\ell} \notag\\
                &=&0;    \notag\\
J_{\beta,b_{\ell}} &=&   J_{b_{\ell},\beta} = 0; \notag\\
J_{b_{\ell_1},b_{\ell_2}} &=&J_{b_{\ell_2},b_{\ell_1}} = 2{\mathfrak Re} \left[\left(\frac{\partial {\bf s}}{\partial b_{\ell_1}} \otimes {\bf a}\right)^H
{\bf K}^{-1}
                \left(\frac{\partial {\bf s}}{\partial b_{\ell_2}} \otimes {\bf a}\right)\right] \notag \\
                &=& \frac{2 {\bf a}^H {\bf a}}{\sigma^2} T^{(\ell_1+\ell_2)}_s \sum^N_{n=1} n^{\ell_1+\ell_2} \notag\\
                &=& \frac{4}{\sigma^2} T^{(\ell_1+\ell_2)}_s \sum^N_{n=1} n^{(\ell_1+\ell_2)},\hspace{0.1in} \ell_1, \ell_2=0,1,\cdots, q. \notag
\end{eqnarray}

The FIM can be re-expressed as:
\begin{eqnarray} \label{eq:FIM}
{\bf J} =& \left[\begin{array}{cccccc}
\frac{2N}{\sigma^2} & 0  & 0 & 0 & \cdots & 0   \\
0 & \frac{2N \sin^2\alpha}{\sigma^2} &0   & 0& \cdots & 0  \\
0 & 0  & {J}_{b_0,b_0}  & {J}_{b_0,b_1} & \cdots & {J}_{b_0,b_q}   \\
0 & 0  & {J}_{b_1,b_0} & {J}_{b_1,b_1} & \cdots & {J}_{b_1,b_q}   \\
\vdots & \vdots & \vdots & \vdots  &\ddots &\vdots \\
0 & 0  & {J}_{b_q,b_0} &  {J}_{b_q,b_1} &\cdots & {J}_{b_q,b_q}   \\
\end{array}\right].
\end{eqnarray}

It is worth noting that both $J_{\alpha,\alpha}$ and $J_{\beta,\beta}$ are decoupled from the other parameters. It follows that the Cram\'{e}r-Rao bounds of $\alpha, \beta$ are {\em independent} of:
\begin{enumerate}
\item[(i)] The polynomial-coefficients $\{b_0,b_1,b_2,\cdots, b_q\}$;
\item[(ii)] The degree of the polynomial-phase signal;
\item[(iii)] The azimuth angle $\beta$.
\end{enumerate}

Therefore, the Cram\'{e}r-Rao bounds of direction-of-arrival are:
\begin{eqnarray}
\mbox{CRB}(\alpha)   &=& \frac{\sigma^2}{2N}, \\
\mbox{CRB}(\beta)   &=& \frac{\sigma^2}{2N \sin^2\alpha}.
\end{eqnarray}
This may seem initially surprising but is in fact
reasonable:
\begin{enumerate}
\item[(a)]
The acoustic vector sensor's array-manifold in (\ref{eq:3+1}) is {\em independent} of the incident source's frequency-spectrum.
Thus, the Cram\'{e}r-Rao bounds will share the same value for the signals with different frequency-spectrums.
\item[(b)]
From (\ref{eq:FIM}), both $J_{\alpha,\alpha}$ and $J_{\beta,\beta}$ are decoupled from the other parameters.
\end{enumerate}
These results are consistent with the studies reported in \cite{NehoraiSPT0994,TichavskySPT1101,TamSJ0809}.

\section{Monte Carlo Simulation}
\label{Sec:simm}
\begin{subfigures}
\begin{figure}
\centering
\begin{minipage}{3in}
\includegraphics[height=6cm,width=8.0cm]{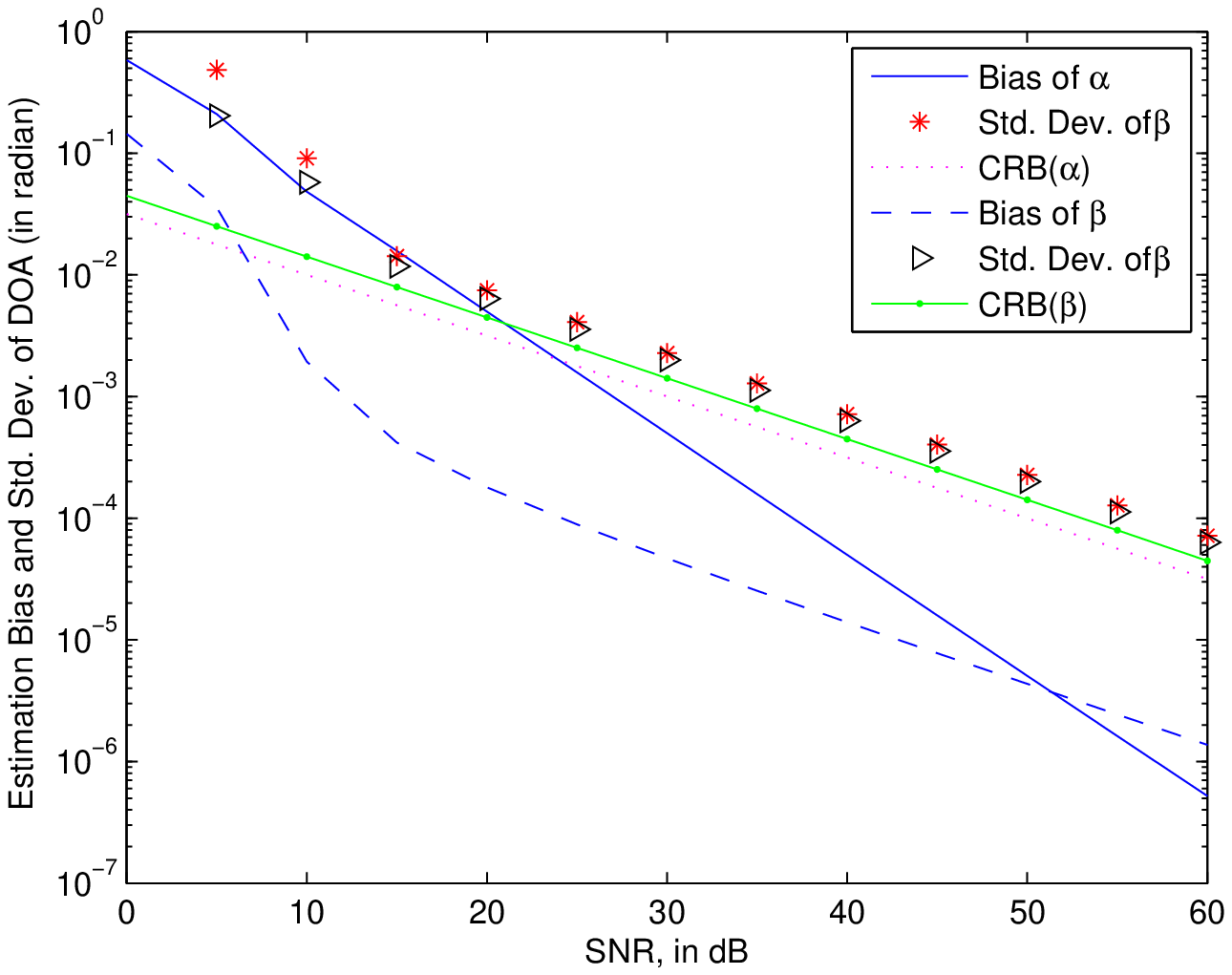}
\caption{Estimation bias and standard deviations of $\{\alpha,\beta\}$ with a $2$-order PPS versus SNR.}
\label{dcbiasstd2}
\end{minipage}
    \hfill
\begin{minipage}{3in}
\includegraphics[height=6cm,width=8.0cm]{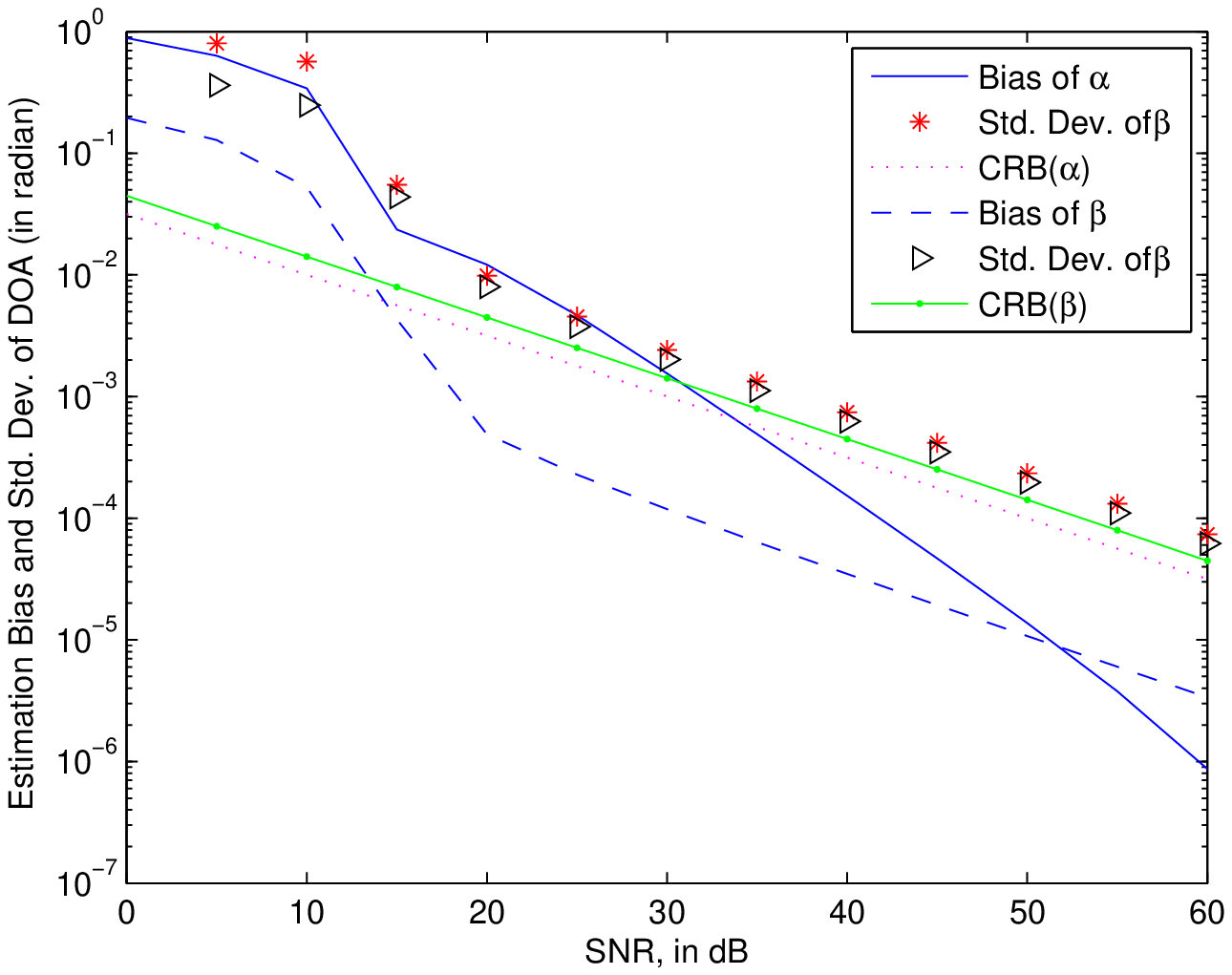}
\caption{Estimation bias and standard deviations of $\{\alpha,\beta\}$ with a $4$-order PPS versus SNR.}
\label{dcbiasstd4}
\end{minipage}
\label{doa}
\end{figure}
\end{subfigures}

\subsection{Examples for Direction Finding}
A $2$-order unity power polynomial-phase signal (a.k.a. LFM or Chirp signal) with $\{b_0=0.05, b_0=0.1, b_2=0.13\}$
is used in this example.
The direction-of-arrival of the source is $\{\alpha,\beta\}=\{45^\circ,60^\circ\}$.
Figure \ref{dcbiasstd2} plots the estimation bias and standard deviations of DOA $\{\alpha,\beta\}$ versus signal-to-noise ratio (SNR) ($1/\sigma^2$).
$1000$ trials are used for each data point on each graph and these estimates use $500$ temporal snapshots.
When SNR$\ge 15$dB, the standard deviations are very close to Cram\'{e}r-Rao lower bounds.
When the SNR is low (SNR$\le 10$dB), the noise affects the algorithm seriously, thus there is a visible gap between the standard deviations and the Cram\'{e}r-Rao bounds.
This is because the {\em multiplicative} noise is introduced when equation (\ref{eq:xqt}) is used to derive the data set ${\bf Y}$.
But when SNR$\ge 15$dB, the noise effect decreases, hence the estimation standard deviations decrease paralleling with the Cram\'{e}r-Rao bounds when the SNR increases.
Figure \ref{dcbiasstd4} plots the estimation bias and standard deviations of $\{\alpha,\beta\}$ in a
$4$-order polynomial-phase signal scenario with $\{b_0=0.05, b_0=0.1, b_2=0.13, b_3=0.23, b_4=0.29\}$.

\begin{subfigures}
\begin{figure*}
\centering
\begin{minipage}{3.2in}
\center
\includegraphics[height=6cm,width=8.0cm]{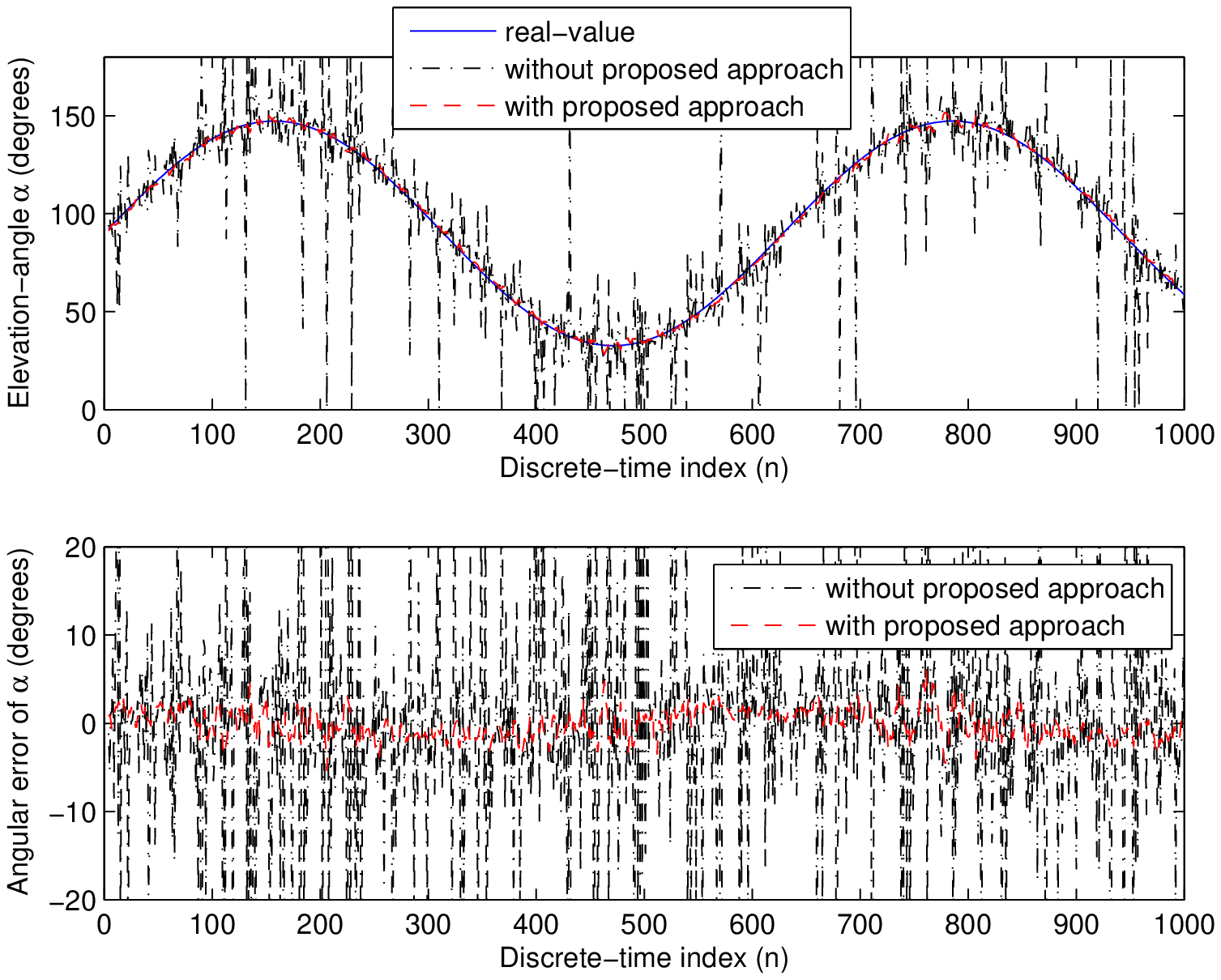}
\caption{Single-forgetting-factor tracking ($\lambda=0.7$) and angular error of the elevation-angle for a PPS source. `without proposed approach' means using the method in \cite{WongKT_OCEANS-AP06} directly, and `with proposed approach' denotes incorporating the proposed pre-processing technique in Section \ref{Sec:ST}.}
\label{theta_sig}
\end{minipage}
    \hfill
\begin{minipage}{3.2in}
\center
\includegraphics[height=6cm,width=8.0cm]{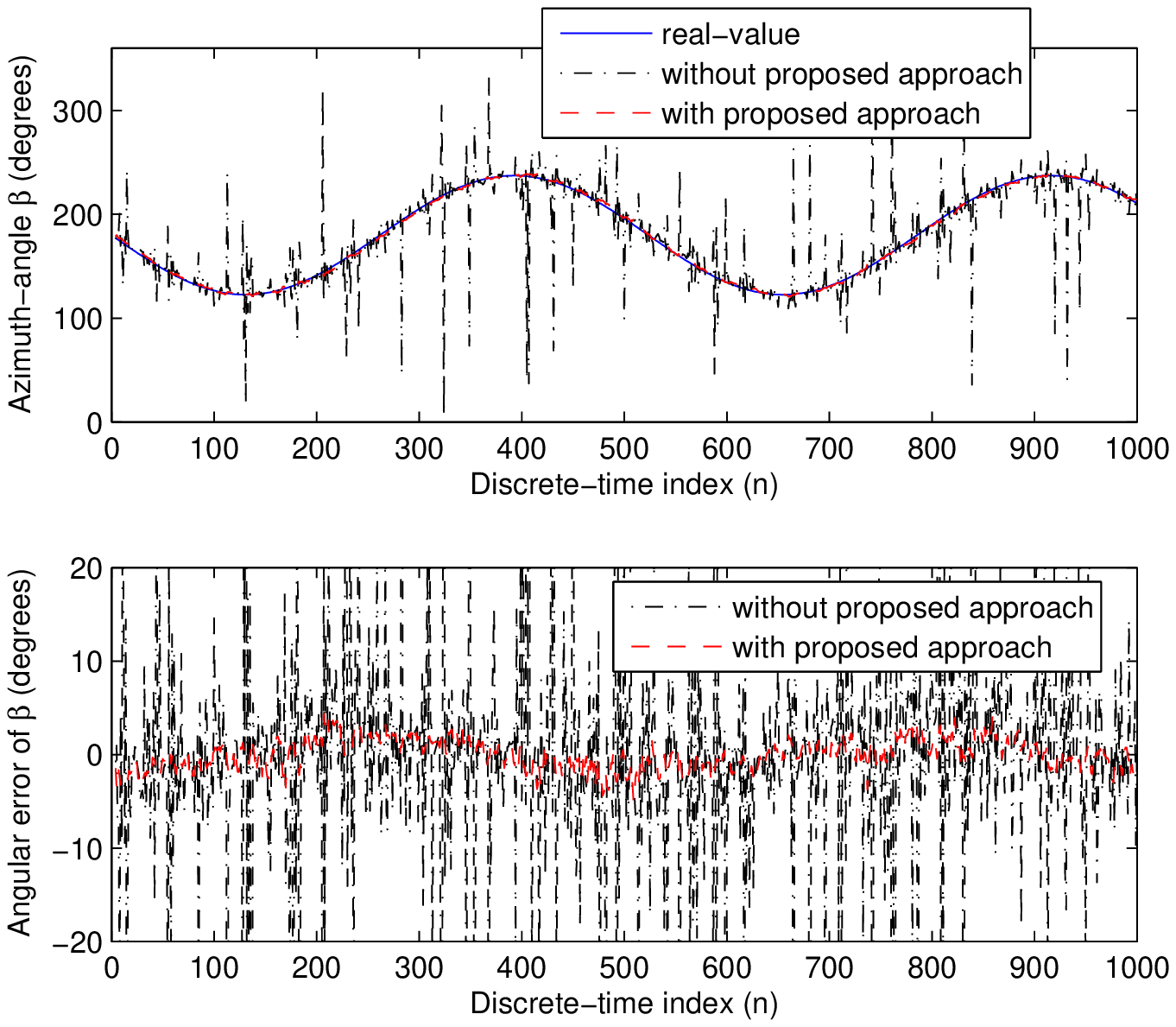}
\caption{Single-forgetting-factor tracking and angular error of the azimuth-angle for a PPS source, same setting as in Figure \ref{theta_sig}.}
\label{phi_sig}
\end{minipage}
\begin{minipage}{3.2in}
\center
\includegraphics[height=6cm,width=8.0cm]{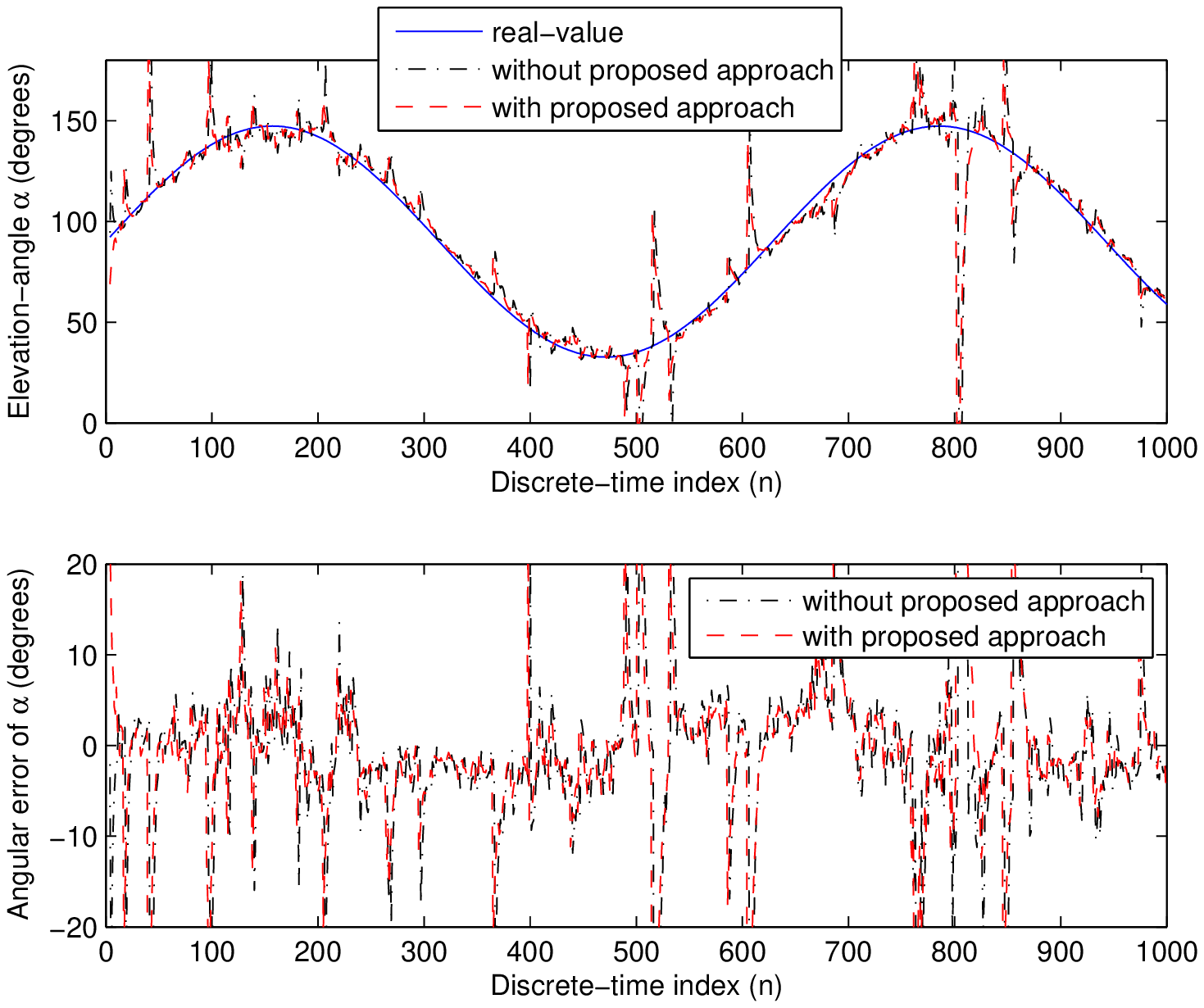}
\caption{Multiple-forgetting-factor tracking ($\lambda_1=0.9,\lambda_2=0.8,\lambda_3=0.7$) and angular error of the elevation-angle for a PPS source.}
\label{theta_mu}
\end{minipage}
\hfill
\begin{minipage}{3.2in}
\center
\includegraphics[height=6cm,width=8.0cm]{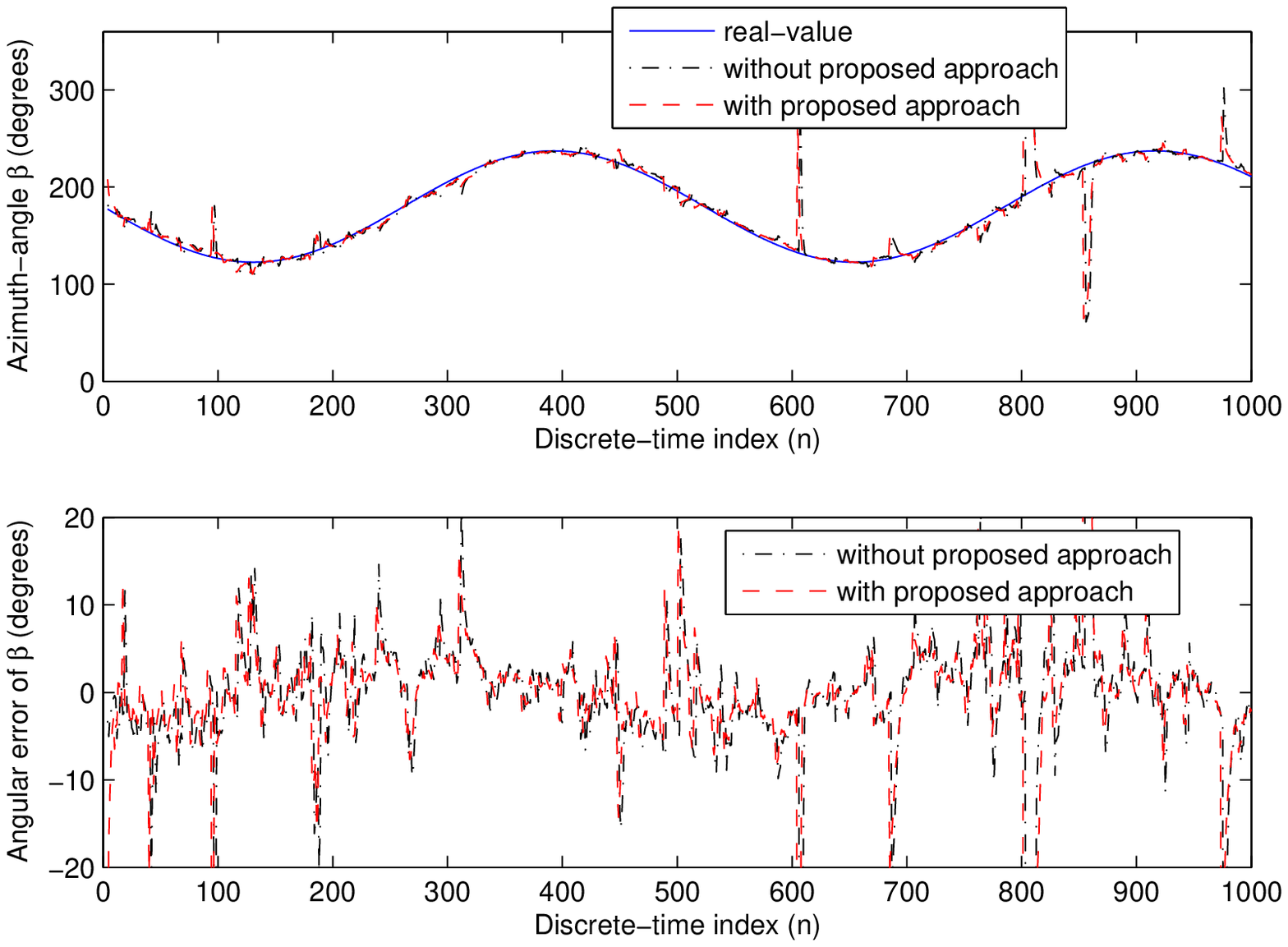}
\caption{Multiple-forgetting-factor tracking and angular error of the azimuth-angle for a PPS source, same setting as in Figure \ref{theta_mu}.}
\label{phi_mu}
\end{minipage}
\label{Strack}
\end{figure*}
\end{subfigures}

\subsection{Examples for Source Tracking}
\begin{table*}
\caption{Angular Error and Standard Deviations of Source Tracking (in degree)}
\centering
\begin{tabular}{|l|ccc|cccc|}
\hline
& $\lambda_1$ & $\lambda_2$ & $\lambda_3$ & Mean of $\alpha_r$ & Std. Dev. of $\alpha_r$ & Mean of $\beta_r$ & Std. Dev. of $\beta_r$ \\
\hline \hline
MFF with{\em out} the proposed technique & $0.9$ & $0.8$ & $0.7$ & $-1.021$ & $11.47$ & $0.459$ & $16.09$\\
\hline
MFF with the proposed technique & $0.9$ & $0.8$ & $0.7$ & $-0.975$ & $11.18$ & $0.700$ & $15.36$\\
\hline \hline
SFF with{\em out}  the proposed technique & $0.9$ & $-$ & $-$ & $-0.478$ & $21.55$ &$-1.561$ & $23.18$\\
\hline
SFF with the proposed technique & $0.9$ & $-$ & $-$ & $-0.233$ & $3.742$ & $0.200$ & $4.065$\\
\hline
SFF with{\em out}  the proposed technique & $0.8$ & $-$ & $-$ & $-0.376$ & $24.45$ & $-0.433$ & $28.00$\\
\hline
SFF with the proposed technique & $0.8$ & $-$ & $-$ & $-0.134$ & $1.853$ & $\color{red}{0.071}$ & $1.930$\\
\hline
SFF with{\em out}  the proposed technique & $0.7$ & $-$ & $-$ & $0.497$ & $20.89$ & $-0.123$ & $28.17$\\
\hline
SFF with the proposed technique & $0.7$ & $-$ & $-$ & $\color{red}{-0.089}$ &$\color{red}{1.689}$ & $0.187$ & $\color{red}{1.570}$\\
\hline
\end{tabular}
\label{tab:ae}
\end{table*}

The time-varying elevation-azimuth angle of the $2$-order moving polynomial-phase signal is modeled as:
\begin{eqnarray}
\alpha(nT_s)&=&\alpha_0 + \sin(\omega_{\alpha} nT_s), \\
\beta(nT_s)  &=&\beta_0 + \sin(\omega_{\beta} n T_s),
\end{eqnarray}
with $(\alpha_0,\beta_0)=(90^\circ, 180^\circ)$, $(\omega_{\alpha},\omega_{\beta})=(0.01,-0.012)$, and $n=1,2,\cdots,1000$.
Figures \ref{theta_sig}-\ref{phi_sig} plot the loci of the source's elevation-angle and azimuth-angle with the single-forgetting-factor (SFF) tracking algorithm ($\lambda=0.7$). The angular errors of $\{\alpha,\beta\}$ are also plotted.
Figures \ref{theta_mu}-\ref{phi_mu} plot the loci of the source's elevation-azimuth angle and the angular errors with the multiple-forgetting-factor (MFF) tracking algorithm ($\lambda_1=0.9, \lambda_2=0.8, \lambda_3=0.7$).
Both the results with and without the proposed pre-processing technique are presented in these figures.
Table \ref{tab:ae} summarizes the angular errors and standard deviations of elevation-angle $\alpha_r$ and azimuth-angle $\beta_r$ for source tracking with different methods.
Qualitative observations obtained from Table \ref{tab:ae} are listed below:
\begin{enumerate}
\item[\{1\}] Both the performances of SFF and MFF methods incorporating the proposed technique in Section \ref{Sec:ST} are better than their counterparts without incorporating the proposed technique.
This can be seen from the standard deviations of $(\alpha_r,\beta_r)$.
\item[\{2\}] The performance of SFF  method incorporating the proposed technique improves significantly in a wide region of $\lambda$ compared with its counterpart without incorporating the proposed technique.
      The result is even better than the MFF method, both with and without the proposed technique.
\item[\{3\}] For the SFF algorithm incorporating the proposed technique, the standard deviations of $(\alpha_r,\beta_r)$ decline when $\lambda$ increases.
\item[\{4\}] The performance of MFF method without the proposed technique is better than the performance of SFF method without the proposed technique.
    This is expected and consistent with the results reported in \cite{WongKT_OCEANS-AP06}.
\end{enumerate}
From the simulation results above, it can be seen that with the proposed technique, the source tracking performance can improve substantially with less computation workload because the performance of the SFF approach outperforms the performances of the other methods.
For the comparison of the computation workload between the SFF and the MFF methods, please refer to \cite{NehoraiSPT1099}.

It is worth pointing out that the SFF and MFF algorithms can be used for any kind of signal model, but the proposed pre-processing technique can only be used in a polynomial-phase source scenario as discussed in this paper.

\section{Conclusion}
An ESPRIT-based algorithm for azimuth-elevation direction-finding of one broadband polynomial-phase signal with an arbitrary degree in investigated is this paper using a single acoustic vector sensor.
The matrix-pencil pair used in the ESPRIT algorithm is derived from the temporally displaced data sets collected by the vector sensor.
Closed-form estimates of DOA are obtained from the eigenvector of signal-subspace.
The proposed algorithm requires neither a priori knowledge of the polynomial-phase signal's coefficients nor a priori knowledge of the polynomial-phase signal's frequency-spectrum.
This is the first time in the literature to use an acoustic vector sensor to resolve the direction-of-arrival of a polynomial-phase signal.
The adaptive tracking algorithms of both the single-forgetting-factor approach and the multiple-forgetting-factor approach are also adapted to incorporate the proposed pre-precessing technique to track a polynomial-phase signal utilizing one acoustic vector sensor.
From the simulation results, the single-forgetting-factor approach with the proposed pre-processing technique can afford better performance than its counterpart without pre-processing technique, and can even offer better performance than the multiple-forgetting-factor approach.
Thus, the proposed source-tracking algorithm can provide novel performance with low computation workload.

\end{document}